\documentclass{aa}
\usepackage{graphicx}
\usepackage{txfonts}
\def\ltsima{$\; \buildrel < \over \sim \;$}
\def\lsim{\lower.5ex\hbox{\ltsima}}
\def\gtsima{$\; \buildrel > \over \sim \;$}
\def\gsim{\lower.5ex\hbox{\gtsima}}

\newcommand{\be}{\begin{equation}}
\newcommand{\en}{\end{equation}}
\newcommand{\ergs}{\rm \ erg \; s^{-1}}

\def\cmdue {\rm \ cm^{-2}}

\def\msole {~M_{\odot}}

\def\deg {^\circ}

\begin{document} 
\title{The X--ray afterglow of the short gamma ray burst 050724}

\author{
S.~Campana\inst{1},
G. Tagliaferri\inst{1},
D. Lazzati\inst{2},
G. Chincarini\inst{1,3},
S. Covino\inst{1},
K. Page\inst{4}
P. Romano\inst{1},
A. Moretti\inst{1},
G. Cusumano\inst{5},
V. Mangano\inst{5},
T. Mineo\inst{5},
V. La Parola\inst{5},
P. Giommi\inst{6},
M. Perri\inst{6},
M. Capalbi\inst{6},
B. Zhang\inst{7},
S. Barthelmy\inst{8},
J. Cummings\inst{8},
T. Sakamoto\inst{8},
D.N. Burrows\inst{9},
J.A. Kennea\inst{9},
J.A. Nousek\inst{9},
J.P. Osborne\inst{4},
P.T. O'Brien\inst{4},
O. Godet\inst{4},
N. Gehrels\inst{8}
}

\offprints{S. Campana, campana@merate.mi.astro.it}

\institute{INAF - Osservatorio Astronomico di Brera, Via Bianchi 46, I-23807
Merate (LC), Italy  
\and
JILA, Campus Box 440, University of Colorado, Boulder, CO 80309-0440, USA
\and
Universit\`a degli studi di Milano-Bicocca, Dipartimento di Fisica, piazza
delle Scienze 3, I-20126 Milano, Italy 
\and
X--Ray and Observational Astronomy Group, Department of Physics and Astronomy,
University of Leicester, LE1 7RH, UK 
\and
INAF, Istituto di Astrofisica Spaziale e Fisica Cosmica di Palermo, via U. La
Malfa 153, I-90146 Palermo, Italy
\and
ASI Science Data Center, via G. Galilei, I-00044 Frascati (Roma), Italy 
\and
Department of Physics, University of Nevada, Box 454002, Las Vegas, NV
89154-4002, USA
\and
NASA/Goddard Space Flight Center, Greenbelt Road, Greenbelt, MD20771, USA 
\and
Department of Astronomy and Astrophysics, Pennsylvania State University, 525
Davey Lab, University Park, PA 16802, USA
}

\date{Received; accepted}

\abstract{Short duration ($\lsim 2$ s) Gamma--ray bursts (GRBs) have been a
mystery since their discovery. Until May 2005 very little was known about
short GRBs, but this situation has changed rapidly in the last few months
since the Swift and HETE-2 satellites have made it possible to discover X--ray
and optical counterparts to these sources. Positional associations indicate
that short GRBs arise in close-by galaxies ($z<0.7$). Here we report on a
detailed study of the short GRB 050724 X--ray afterglow. This burst shows
strong flaring variability in the X--ray band. It clearly confirms early
suggestions of X--ray activity in the 50--100 s time interval following
the GRB onset seen with BATSE. Late flare activity is also observed. These
observations support the idea that flares are related to the inner engine for
short GRBs, as well as long GRBs.   
}

\titlerunning{The afterglow of the short GRB 050724}
\authorrunning{S. Campana et al.}

\maketitle

\section{Introduction}
\label{intro}

Gamma Ray Bursts (GRBs) are intense flashes of $\gamma$--ray radiation
outshining all other sources in the gamma--ray sky.  GRBs can be divided into (at
least) two classes based on their temporal and spectral properties: short and
long GRBs. Short GRBs have a typical duration of 0.2 s and can last from a few
milliseconds to several seconds. Short GRBs are harder (i.e. have more
flux at higher energies) than long bursts. Thus, in a plane [$T_{90}$ --
hardness ratio] they lie in a region mostly distinct from long GRBs
(Kouveliotou et al. 1993). 
A more general definition of short burst has been given by Norris \& Bonnell
(2006) that the initial spike exhibits negligible spectral evolution at
energies above $\sim 25$ keV. Short GRBs comprise about $25-30\%$ of the BATSE
(25--350 keV) sample and about $12\pm4\%$ of the Swift sample. This lower
fraction is due to the softer energy band (15--150 keV) of the Burst Alert
Telescope (BAT, Barthelmy et al. 2005a) onboard Swift (Gehrels et al. 2004).

Despite the great improvements in our knowledge of long GRBs in the last
decade, very little was known on short GRBs. This situation changed
drastically starting from May 2005, when the Swift satellite (Gehrels et
al. 2004) was able to detect and accurately localize for the first time the
X--ray afterglow of the short burst GRB 050509B (Gehrels et al. 2005). 
This burst lies near a luminous, non star-forming ($<0.1\msole$ yr$^{-1}$)
elliptical galaxy at $z=0.22$ located in a cluster of galaxies. The {\it a
posteriori} probability of the position being close to such a nearby, luminous
galaxy by coincidence is $\sim 10^{-4}$ (fainter objects in the error circle 
have been reported by Bloom et al. 2005). Two months later the HETE-2 satellite
localized GRB 050709, the first short GRB with an optical counterpart. The
object was coincident with a weak X--ray source and was located inside a
galaxy at $z=0.16$ (Covino et al. 2005a; Hjorth et al. 2005; Fox et al. 2005).  
Spectroscopic observations showed that the dominant stellar population is
relatively young ($\sim 1$ Gyr) and that there is ongoing star formation
($\sim 0.2\msole$  yr$^{-1}$, Fox et al. 2005, Covino et al. 2005a). 
GRB 050724 was the first short GRB with low energy prompt X--ray emission
lasting for 100 s after the main short pulse, strong early X--ray afterglow
and an unusual X--ray re-brightening at $3\times 10^4$ s (a short account of
these data has been given by Barthelmy et al. 2005b). The X--ray and optical
afterglow of this burst (D'Avanzo et al. 2005; Gal-Yam et al. 2005; Barthelmy
et al. 2005b; see also Covino et al. 2005b) is located off-center in an
elliptical galaxy at  $z=0.258$ with a very low star-formation ($<0.02\msole$
yr$^{-1}$; Prochaska et al. 2005a). The X--ray afterglow of the short GRB
050813 simply faded without any flare, leading to a localization consistent
with a distant ($z\sim 0.7$) cluster of galaxies (Gladders et al. 2005).
The host galaxy for this burst has not been found yet (Prochaska et al. 2005b;
see Moretti et al. 2005, for an updated position).
In addition, GRB 050724 is the first and only GRB for which a radio afterglow
has been detected (Soderberg, Cameron \& Frail 2005).

In this paper we present a thorough analysis of the X--ray properties of
GRB 050724 as observed by Swift. In Section 2 we present the data and in
Section 3 their spectral and temporal analysis. Section 4 is dedicated to 
the discussion and Section 5 to the conclusions.

\section{Swift data}
\label{data}

GRB 050724 (Covino et al. 2005b) was discovered at 2005-07-24 12:34:09 UT by 
the BAT. Swift slewed to the
burst position in only 74 s. The X--Ray Telescope (XRT, Burrows et al. 2005a)
observations started at 12:35:22.9 UT in Auto State. XRT detected a rapidly
fading source (Romano et al. 2005a). The UltraViolet and Optical Telescope
(UVOT, Roming et al. 2005) did not detect the afterglow ($V>18.8$, Chester et
al. 2005). 

The XRT observed the GRB position nine times. The log of these observations is
presented in Table 1. The first observation is split into two parts with
two different observing modes.
For bursts observed in Auto State the XRT observations start in
Window Timing (WT) mode (providing just 1D imaging) and when the source gets
below a predefined count rate threshold XRT switches automatically to the more
common Photon Counting mode (PC, providing 2D imaging and full
spectroscopic resolution; for a
description of XRT observing modes, see Hill et al. 2004).
Cross-calibration between modes assures that the two modes, PC and WT, provide
the same rate (within a few percent) on steady sources. 

GRB 050724 was detected in the first two observations and in the
combination of observation 7 and 9 (47108 s). In the following, we do not
consider the other observations.
Finally, we note that a dust-scattered X--ray halo has been discovered around
GRB 050724 (Romano et al. 2005b). This topic has been fully covered by
Vaughan et al. (2005) and it will not be discussed in the present work.

\begin{table}
\label{datalog}
\caption{Observation log.}
\begin{tabular}{ccccc}
\hline
OBS.ID.    & Start time&Mode& Exp. time & Count rate  \\
           & (s)$^*$   &    & (s)       & $3\,\sigma$ UL\\
\hline
00147478000& 74	       & WT & 8438      & \\
           & 189       & PC & 27737     & \\
00147478001& 127716    & PC & 16947     & \\ 
00147478002& 215626    & PC & 135       & $<1.1\times 10^{-1}$ \\
00147478003& 215679    & PC & 2121      & $<6.2\times 10^{-3}$ \\ 
00147478004& 377054    & PC & 5244      & $<2.8\times 10^{-1}$ \\ 
00147478005& 377083    & PC & 2218      & $<5.8\times 10^{-3}$ \\ 
00147478006& 388061    & PC & 267       & $<1.6              $ \\ 
00147478007$^+$& 388119    & PC & 25979     & $<1.0\times 10^{-3}$ \\ 
00147478008& 474403    & PC & 1383      & $<1.1\times 10^{-2}$ \\ 
00147478009$^+$& 474582    & PC & 21129     & $<9.2\times 10^{-4}$ \\ 
\hline
\end{tabular}

$^*$ Time from the BAT trigger time.

$^+$ These two observations when summed together lead to a (low significance)
detection at a rate of $(2.6\pm1.1)\times 10^{-4}$ c s$^{-1}$.

\end{table}

\begin{figure*}
\centering
\includegraphics[height=16cm,angle=-90]{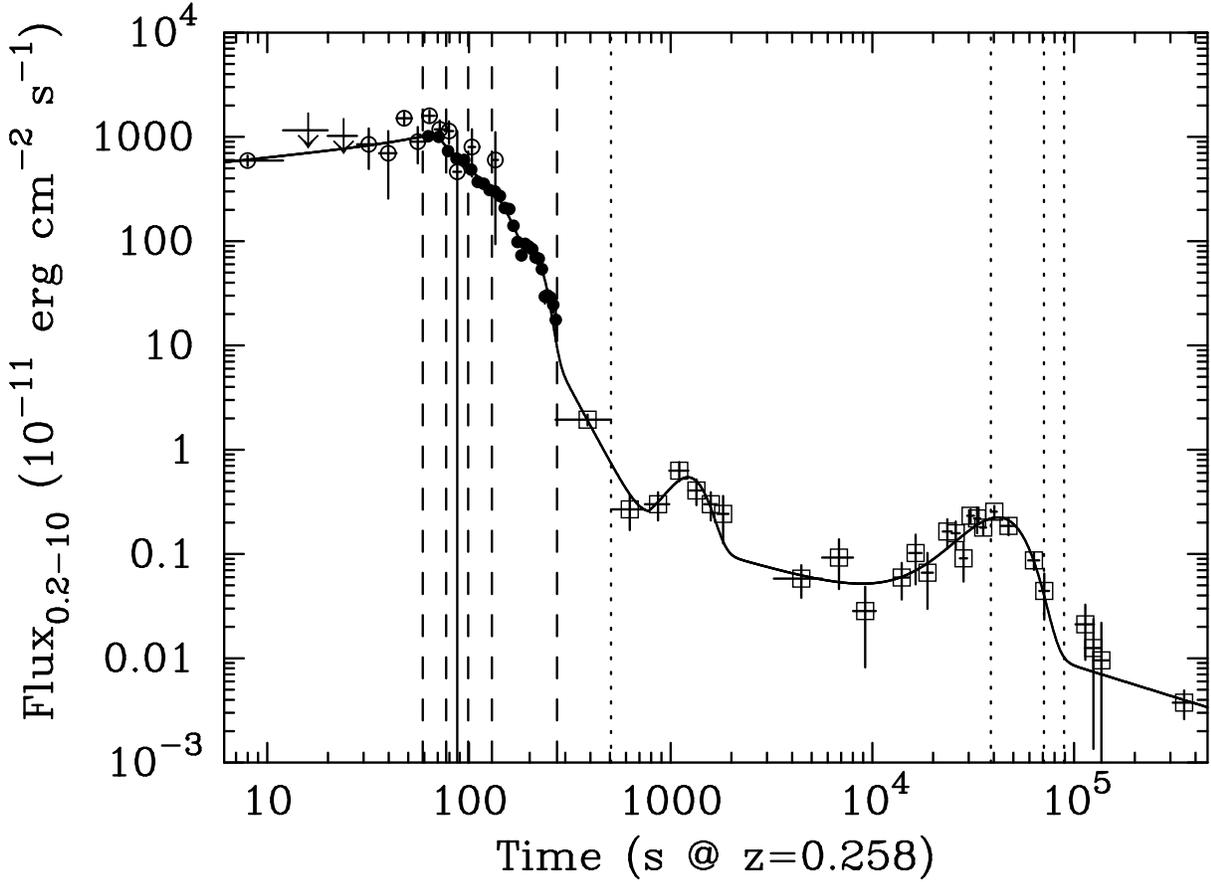}
\caption{X--ray light curve of GRB 050724. Count rates have been converted
into 0.2--10 keV unabsorbed flux after spectral modelling (see text). Times
have been converted to the GRB progenitor reference time. Open circles mark
BAT data (as well as $3\,\sigma$ upper limits) extrapolated to the
0.2--10 keV energy band; filled circles mark XRT-WT
data and open squares XRT-PC data, respectively. The continuous line
represents the best fit model obtained with four Gaussians and a triple power law (see
text). Dashed lines indicate WT mode time intervals selected for spectral
analysis. Dotted lines indicate time intervals used for spectral analysis in
PC mode. 
}
\label{lc}
\end{figure*}

\section{Data analysis}
\label{analysis}

\subsection{Burst Alert Telescope}
The BAT light curve shows a short bright peak with 0.25 s duration.  There is
additional prompt emission in the time interval [-20, 20] s with a $T_{90}$ of
$3.0\pm1.0$ s (Barthelmy et al. 2005b). Besides the short
spike, there is an extended low-flux tail lasting for $\gsim 200$ s (Krimm et
al. 2005). If this extended tail is included, the duration becomes
$T_{90}^*=152.4\pm9.2$ s. 
Since this $T_{90}$ is longer than the canonical 2 s, a more general
definition of short burst has been suggested by Norris \& Bonnell 
(2006) based on the unique properties of the initial spike: this spike should
exhibit negligible spectral evolution at energies above $\sim 25$ keV.
The first peak is relatively hard with a power law
photon index $\Gamma=1.71\pm0.16$ ($90\%$ confidence level, throughout the
paper), after which the emission softens to $\Gamma=2.5\pm0.2$. The peak
15--350 keV fluence (1 s) is $(2.6\pm0.3)\times 10^{-7}$ erg cm$^{-2}$ 
(in the $T_{90}$ is $(3.9\pm1.0)\times 10^{-7}$ erg cm$^{-2}$)
whereas the total fluence is $(1.5\pm0.3)\times 10^{-6}$ erg cm$^{-2}$ (see
also Barthelmy et al. 2005b). From the BAT analysis it is clear that the bulk
of the energy is not emitted in the short initial spike ($<1$ s or $T_{90}$)
but in the extended tail. Moreover, the bump at 50--100 s is consistent in
time and spectrum with the soft excess observed with BATSE after summing a
large number of bright short bursts (Lazzati et al. 2001). 

\subsection{X--Ray Telescope}

All data were processed with the standard XRT pipeline within  
FTOOLS 6.0 ({\tt xrtpipeline} v. 0.8.8) in order to produce screened event
files. WT data were extracted in the 0.5--10 keV energy range, PC data in the
0.3--10 keV range. Standard grade filtering was adopted (0--2 for WT and
0--12 for PC, according to XRT nomenclature, see Burrows et al. 2005a). From
these data we extracted spectra and light curves using regions selected to
avoid pile up and to maximize the signal to noise (see Table 1 and below). In
WT mode we adopted the smaller 
than standard extraction region of $20\times 20$ pixels ($47''\times
47''$) along the WT line in order to avoid any small contribution from the
dust-scattered halo (Vaughan et al. 2005). In PC we used a small circular
extraction region of 10 arcsec. Appropriate ancillary response  files were
generated with the task {\tt xrtmkarf}, accounting for Point Spread Function
(PSF) corrections. The latest response matrices (v.007) were used. 

\subsection{Image analysis}

An initial source position has been reported in Barthelmy et al. (2005b). We
take advantage of the second observation where the pile-up is completely
absent and there is no contamination from the dust-scattered halo. Using the
FTOOLS task {\tt xrtcentroid}, which includes also systematic effects, we
derive RA(J2000)=16$^{\rm h}$ 24$^{\rm h}$ 44.40$^{\rm s}$,
DEC(J2000)=--27$^{\rm o}$ $32'$ $27.4''$ with a $90\%$ error radius of
$4.2''$. This includes a  correction for the satellite boresight (Moretti et
al. 2005). This improved XRT position is $1''$ from the previous determination
(Barthelmy et al. 2005b) and $0.5''$ and $0.4''$ from the Chandra and optical
(VLT, D'Avanzo et al. 2005; Barthelmy et al. 2005b), and VLA positions
(Soderberg 2005), respectively.  

\subsection{Timing analysis}

The 0.2--10 keV light curve of the GRB 050724 afterglow is shown in
Fig. \ref{lc}. The light curve is in the GRB rest-frame and the effective
energy range is 0.25--12.6 keV. The early part of the curve comes from the BAT
data extrapolated to the XRT energy band with the BAT spectral model described
above. The second part is from the XRT data. Fluxes 
have been computed according to the best fit models (see below). It is clear
from Fig. \ref{lc} that the light curve is extremely rich.  
There is a steep decay following the GRB proper (possibly
with some structure overlayed) and two big flares. We fit this light
curve with a doubly-broken power law (as usually done for long GRBs as well,
see Chincarini et al. 2005; Nousek et al. 2005; O'Brien et al. 2005) plus four
Gaussians. Formally this model does not provide a good representation of the
data since we have  a reduced $\chi^2_{\rm red}=2.5$ (with 47 degrees of
freedom, dof), but it catches the main features of the afterglow light curve
(see Fig. 1, the high $\chi^2$ is mainly due small scale fluctuations). 
The first power law has a rising index $\alpha_1=+0.2\pm0.1$ 
and accounts for the prompt emission observed by the
BAT (lack of data in this first part is due to upper limits in the BAT light
curve). The first time break is at $t_1=69\pm2$ s, indicating when the  
very steep decay starts. The second power law has
$\alpha_2=-3.6\pm0.1$. Then the power law decay flattens to
$\alpha_3=-0.6\pm0.2$ after $t_2=775\pm395$ s. During the first $\sim 1000$ s
the source fades by about four orders of magnitude in flux. 
On top of this behaviour we have to include at least four main (Gaussian)
flares with start time $109\pm7$, $215\pm5$, $1226\pm165$ and $42100\pm2230$
s, respectively\footnote{The statistical significance of the third flare is
about $2.2\,\sigma$ computed in the 300--2000 s time interval}. The widths
(Gaussian $\sigma$) of the four flares are $40\pm8$ s, $25\pm6$ s, $263\pm194$
and $14590\pm2425$ s, respectively ($\delta t/t$ of the flares are
$0.37\pm0.08$, $0.12\pm0.03$, $0.21\pm0.16$ and $0.35\pm0.02$). Major
Gaussian-like flares have been reported in GRB 050502B (Burrows et al. 2005b;
Falcone et al. 2005) and XRF 050406 (Romano et al. 2005c).
We also investigate different models for the first two flares. An exponential
tail or a Lorentzian model provide worse fits than a Gaussian model. After
MacFadyen et al. (2005) we also consider a fast 
rise exponential decay (FRED) for the first two flares\footnote{The FRED
flares have been modelled as $x^r\times\exp{-(t-t_0)/t_c}$, where $t$ is the
time, $t_0$ is the start time, $t_c$ is decay time and
$x=2.71\,(t-t_0)/(r\,t_c)$, accounts for the rise. The peak in this
parametrization is at $r\,t_c-t_0$.}. The fit is slightly improved
($\chi^2_{\rm red}=2.2$, with 45 dof, $2.5\,\sigma$ according to an F-test).
The peaks of the two flares in this case move to 112 s and 187 s.

 From the theoretical point of view, flares are sometimes modelled with power
laws (e.g. Zhang et al. 2005). We tried smoothly joined power laws
to describe the flares on top of the doubly-broken power law. 
The fit provides a slightly better description of the data with $\chi^2_{\rm red}=2.3$
(46 dof). The times of the two breaks are at $t_1=70\pm2$ s and $t_2=820\pm180$
s and the power law indices are $\alpha_1=0.2\pm0.1$, $\alpha_2=-3.6\pm0.2$
and $\alpha_3=-0.6\pm0.2$. Peak time of the four flares are $148\pm7$ s,
$210\pm9$ s, $994\pm106$ and $44478\pm9298$ s (here and in the following times
are rest-frame). The rising exponents (with
reference to the BAT trigger) are not well constrained, the one relative to the
first flare is very flat (fixed to $1$), the following two are very steep
(fixed to $10$) and the last one is $2.2\pm0.7$. The decaying exponents are
better constrained and all very steep: $9\pm3$, $11\pm4$,
$3^{+3}_{-1}$ and $4^{+3}_{-2}$, respectively. A model with four flares
without the underlying broken power law provides a worse fit with $\chi^2_{\rm
red}=3.6$.   

The total fluence of the GRB 050724 afterglow in the 0.2--10 keV energy band
is $1.4\times 10^{-6}$ erg$\cmdue$ in the $10-10^6$ s time interval. This is
larger than the burst prompt fluence (1 s, extrapolated to the same band) by a
factor of $\sim 10$. The fluence of the first flare is $2.5\times 10^{-7}$
erg$\cmdue$ ($21\%$ of the total fluence). The fluence of the second flare is
$2.8\times 10^{-8}$ erg$\cmdue$ ($2\%$ of the total). Results are similar
(within a few percent of the total fluence) also in the case of a FRED
modelling. The last two flares have $2.8\times 10^{-9}$ erg$\cmdue$ ($0.2\%$)
and $7.7\times 10^{-8}$ erg$\cmdue$ ($7\%$), respectively. Flares account for
$\sim 30\%$ of the total fluence in the 0.2--10 keV energy band.

\begin{figure}
\centering
\includegraphics[height=8cm,angle=-90]{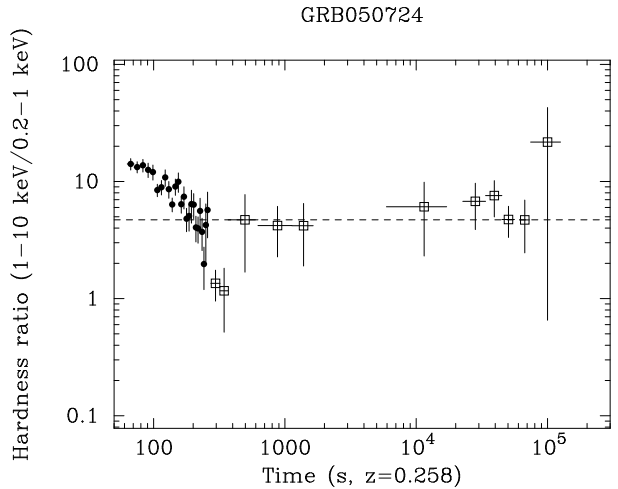}
\caption{Hardness ratio light curve. The ratio is between 1--10 keV to 0.2--1
keV. Time has been corrected for redshift time dilation. Filled circles mark
XRT WT data, open squares XRT PC data, respectively. The dashed line indicates
the mean value of the data starting from 370 s from the burst trigger (rest frame).
}
\label{hardness}
\end{figure}

\subsection{Spectral analysis}

Given the complex behaviour of the light curve we performed the spectral
analysis over different segments of the light curve. A first indication of
spectral changes can be obtained from a band (hardness) ratio analysis (we used the
1--10 keV over 0.2--1 keV energy bands, Fig. \ref{hardness}). It is apparent
that the rapid decay up to 500 s after the burst shows a transition from hard
to soft. After $\sim 500$ s the hardness ratio becomes stable, despite the
presence of the two late flares (rest frame time $t>500$ s). The rise in hardness
ratio after the flare is uncommon: in other GRBs the ratio typically returns
to a low level after each flare.

For a proper spectral fitting, we divided the first part of the XRT light
curve into five intervals containing at least 500 counts. The first four are
in WT mode and the last one is in PC mode. The data were rebinned to have 30
counts per energy bin. The X--ray absorption has been modelled with a
constrained Galactic component ($3.4-4.2\times 10^{21}\cmdue$, Vaughan et
al. 2005), plus an intrinsic ($z=0.258$) column density. The model used within
XSPEC (11.3.2) is {\tt tbabs}. For the spectrum we adopted a simple power
law. 

The overall fitting is good, providing a reduced $\chi^2=1.03$ (321 degrees of
freedom and $36\%$ null hypothesis probability).
The intrinsic column density is constant within the errors.
As can be seen from the hardness ratio plot and from the detailed spectral
analysis (Table 2) there is a clear trend from hard to soft in the first 400 s
(rest frame) after the burst. This is observable in the power law photon index
which is softening from $1.4\pm0.1$ to $3.3\pm0.4$.

We also analyzed the last part of the light curve adopting a
different approach. In this case the number of photons is relatively
low and we fit an absorbed (Galactic plus local) power law using
Cash-statistics (see second part of Table 2). The first PC interval is
at the end of the second part (the steepest) of the 3-power law component and
well after the first two flares. The spectrum during this interval is 
very soft ($\Gamma=3.3\pm0.4$. The remaining three intervals instead have very
similar spectra, somewhat harder than the first PC interval.

Given the complex behaviour of the X--ray afterglow we also tried a cut-off
power law model since this model can accommodate spectral variations as
demonstrated in the case of GRB 050502B's flare (Falcone et al. 2005).
The fit confirms the results above with cut-off energies decreasing with time
(the overall $\chi^2_{\rm red}=1.01$ but the chance improvement computed with
an F-test is just $5\%$).

\begin{table}
\caption{Spectral fits.}
\begin{tabular}{ccccc}
Interval    & $N_H$ @ $z$          &Power law     & Mean flux  \\
(s, rest frame)&($10^{21}$ cm$^{-2}$)&$\Gamma$       & (0.5--10 keV) \\
\hline
59--77      &$2.3^{+2.0}_{-1.2}$&$1.41^{+0.09}_{-0.09}$&$9.3\times 10^{-9}$\\
77--99      &$4.0^{+2.0}_{-1.3}$&$1.84^{+0.11}_{-0.10}$&$6.0\times 10^{-9}$ \\
99--130     &$2.3^{+1.8}_{-1.1}$&$1.91^{+0.11}_{-0.10}$&$4.7\times 10^{-9}$\\
130--268    &$1.9^{+1.5}_{-0.8}$&$2.34^{+0.10}_{-0.09}$&$8.1\times 10^{-10}$\\
268--89232  &$<0.9^\#$         &$1.98^{+0.16}_{-0.20}$&$1.4\times 10^{-12}$\\
\hline
268--506$^*$& $<0.7^+$         &$3.30^{+0.41}_{-0.36}$&$2.3\times 10^{-11}$\\
507--38647  &                &$1.76^{+0.16}_{-0.17}$&$1.9\times 10^{-12}$\\
41152--70835&                &$1.72^{+0.21}_{-0.21}$&$2.1\times 10^{-12}$\\
73793--89232&                &$1.78^{+0.70}_{-0.68}$&$6.6\times 10^{-13}$\\
\hline
\end{tabular}

$^\#$ the $3\,\sigma$ ($\Delta\chi=6.43$) upper limit is $2.0\times10^{21}\cmdue$

$^+$ value fixed for all the four PC observations. The $3\,\sigma$ upper limit
is $1.2\times10^{21}\cmdue$.

$^*$ adding a cut-off to the power law, the photon index is constrained to be 
$3.03^{+0.80}_{-2.11}$ and the high energy cut-off $>0.5$ keV.

\end{table}

\section{Discussion}
\label{discu}

GRB 050724 has a rich temporal and spectral phenomenology.
The first peak in the BAT data is hard and typical of the short/hard GRB 
population. Long-soft emission following the hard peak has been
reported by Lazzati et al. (2001), similarly to what is observed
here for GRB 050724. Moreover, GRB 050709 discovered by HETE-2 (Villasenor et 
al. 2005) shows a similar soft bump in the Soft X--ray Camera 
(SXC) and a late flare (Fox et al. 2005).
We can speculate that these bumps are a common characteristic
of at least some of the short GRBs. This is at variance with at least three
other short/hard GRBs: GRB 050509B (Gehrels et al. 2005), GRB 050813 (Retter et
al. 2005) and GRB 050906 (Krimm et al. 2005), that do not show signs of
extended emission (GRB 050509B has only 11 photons collected by XRT within
1000 s). 
These characteristics may suggest two classes of short GRBs: one with
strong soft emission following the hard short peak and the other with
very faint afterglows. We can speculate that this second class might be
related to Soft Gamma--ray Repeaters
(SGR) in closeby galaxies (and in fact the estimated redshift of the galaxies
positionally closer to the these GRBs are lower than the ones showing soft bumps)
even if the total energy must be an order of magnitude larger than the brightest
flare observed to date from SGR 1806--20 (Dec 27, 2004; e.g. Palmer et 
al. 2005). Alternatively, short GRBs may exist with no bumps, as already
observed for long GRBs and SGRs comprising a very small population ($\lsim
1\%$). Weak afterglows might also be produced by a compact object (black hole
- neutron star) mergers with later outbursts due to a longer disk lifetime
(Barthelmy et al. 2005; Page et al. 2005; Davies, Levan \& King 2005) or to
bursts occurring in low density environment (Vietri 1997; Page et al. 2005). 

The most striking feature in the afterglow of GRB 050724 are the big
flares and the extended emission following the short initial spike. The first
two flares have been modelled both as two Gaussians and two FRED-like bursts
(with the largest portion of the energy released in the first one). The total 
(0.2--10 keV) energy of the first flare is $6\times 10^{49}\ergs$. 
If alone it might be consistent with the model by MacFadyen et al. (2005)
suggesting that these flares arise from the interaction of the GRB outflow with
a non-compact stellar companion. However, the presence of multiple bumps (and
in particular the second one) strongly argue against this interpretation. In
addition, these bumps have short durations ($\delta{t}/t<1$), implying that
they originate in a region more compact than the external shock. Delayed
activity from the inner engine is a more plausible explanation (Burrows et
al. 2005b; Romano et al. 2005c; Falcone et al. 2005). 

A different approach (even if less detailed) might rely on
the self-gravitating fragments of (unstable) neutronized matter left over 
during a binary neutron star coalescence due to sausage instability
(Faber et al. 2005; Colpi \& Rasio 1994; see also Perna et
al. 2005). An interesting feature of the first flare is that its tail
is characterized by peculiarly soft emission, softer than the late
time emission dominated by the external shock component. Analogously
to what is observed in long GRBs (Burrows et al. 2005b; Zhang et al. 2005; Lazzati
\& Begelman 2005) the best explanation for the steep decay of
the early afterglow is large angle emission from the internal shock
phase. Such an emission component is expected to be very soft due to
the large redshift implied by the relativistic de-beaming. This is an
additional piece of evidence that the first bump is related to the inner engine
and not to the interaction with a distant companion star.

The last two flares are less energetic ($6\times 10^{47}$ erg and
$2\times 10^{49}$ erg). These two flares
may be the result of the interaction of the first ejecta (the one relative
to the prompt emission of the short GRB), with the delayed ejecta of the two
secondary flares. In particular, we note that the intensity ratio of the two
early flares is similar to that of the two late flares. 

Finally, the absence of any break in the light curve ($t_b>3.5\times
10^5$ s) indicates that we have a large opening angle (although we caution the
reader that the possible presence of flares makes the interpretation of the
light curve in terms of power laws less secure). 
Under the simplifying assumption of a constant
circumburst density medium of number density $n$, a fireball
emitting a fraction of its kinetic energy in the prompt $\gamma-$ray
phase would show a break in its afterglow light curve when its
bulk Lorentz factor $\Gamma_{\rm b}$ becomes of the order of
$\Gamma_{\rm b}=1/\theta$, with $\theta=0.161\,(t_{b,1d}/(1+z))^{3/8}\,(n\,
E_{iso, 52}^{-1})^{1/8}$ (with $E_{iso, 52}$ the
isotropic energy in units of $10^{52}$ erg and $t_{b,1d}$ the break time in
units of 1 d; Rhoads 1997; Sari et al. 1999). 
In the case of GRB 050724 the isotropic total energy is $E_{iso}\sim 10^{51}$
erg (a lower value would increase the angle even more) and for a standard
density of 1 particle per cm$^{-3}$, we obtain $\theta\gsim 20\deg$, a very
large value compared to long GRBs. In order to have a smaller opening angle
($\theta \lsim 10\deg$) we would need either a lower particle density $n\lsim
3\times 10^{-3}$ cm$^{-3}$ (or a break time hidden by the last flare,
i.e. $t_b\lsim70,000$ s).

\section{Conclusion}
\label{conclu}

GRB 050724 is the first short GRB showing strong flaring and extended emission
activity following short initial spike. This flaring activity has been
collectively revealed 
by averaging short GRB light curves detected by BATSE in the 50--100 s time
interval (Lazzati et al. 2001). This indicates {\it a posteriori} that this
extended activity is likely a common characteristic of a good fraction of short
GRBs. Moreover, these observations prove that flaring activity is a
characteristic not just related to long GRBs (e.g. Burrows et al. 2005b) but
to the GRB phenomenon itself.

We observe flaring activity both in the first stages of the X--ray afterglow
(i.e. within a few hundred seconds from the GRB onset) and at late times.
Energetics ($30\%$ of the afterglow fluence) and duration ($\delta t/t <0.4$)
considerations point toward a prolonged engine activity as already invoked for
long GRBs (Burrows et al. 2005b; Falcone et al. 2005; Romano et al. 2005c). 
Finally, the lack of any break in the X--ray light curve likely indicates
either a large beaming angle or a low density medium surrounding the
GRB and, therefore, a different progenitor from long GRBs.

\begin{acknowledgements}
This work is supported at OAB by funding from ASI on grant number
I/R/039/04, at Penn State by NASA contract NAS5-00136 and at the
University of Leicester by the PPARC on grant numbers PPA/G/S/00524
and PPA/Z/S/2003/00507. DL acknowledges support from NSF grant
AST-0307502 and NASA Astrophysical Theory Grant NAG5-12035. We
gratefully acknowledge the contributions of dozens of members of the
Swift team, who helped make this Observatory possible.
\end{acknowledgements}

\end{document}